\input{aipcheck}

\documentclass[
    ,final            
  ]
  {aipproc}

\layoutstyle{8x11single}

\begin{document}

\title{X-Ray Afterglows}


\author{Paul O'Brien}{
  address={Department of Physics \& Astronomy, University of Leicester, UK}
}

\author{Dick Willingale}{}

\begin{abstract}
 
We summarise the X-ray temporal and spectral variability
properties of GRBs as observed using the {\it Swift} satellite.
Despite much individual complexity, the flux and spectral variability
can be reasonably well described by a combination of two components --
which we denote as the prompt and the afterglow. The first, prompt component
consists of the burst and its initial decay while the second, afterglow
component fits the X-ray plateau phase and subsequent decline observed
in the majority of GRBs. When strong spectral variability
occurs it is associated with the prompt component while the X-ray
plateau and later emission shows little if any spectral variability.
We briefly compare the observations with some of the proposed models.
Any model for the early or late emission must explain the differences
in both temporal and spectral behaviour.

\end{abstract}

\maketitle


\section{Introduction}

The {\it Swift} satellite \citep{ge04} is discovering around 100 GRBs
per year, the vast majority of which are detected with the X-ray
Telescope (XRT) \citep{bur05} following a trigger from the Burst Alert
Telescope (BAT) \citep{bart05}.  The behaviour of the X-ray temporal
and spectral data is complex.  In the majority of bursts ($\sim
75$\%), the X-ray flux declines fairly rapidly, sometimes very
rapidly, within the first few thousand seconds (observed time), before
a shallower decay, or plateau, phase occurs which typically last a few
tens of thousands of seconds but can last up to a day
(e.g. \citep{ta05, nou06, ob06, will07}). X-ray flares are also widely
seen ($\approx 50$\% of GRBs), usually during the inital decay up to
the start of the X-ray plateau. The flares have properties consistent
with being due to an internal process
\citep{chin07, falcone07}. After the X-ray plateau, the flux decays at
a rate fairly consistent with the predictions of standard pre-{\it
Swift} GRB afterglow models
\citep{will07}. This light curve behaviour is now often referred to as
the ``canonical X-ray light curve''. However, a significant number of
GRBs decay in a more gradual, fairly uniform manner from the start
(e.g. \citep{mun07}). 

\section{X-ray light curves}

\begin{figure}
  \includegraphics[angle=-90,width=.4\textheight]{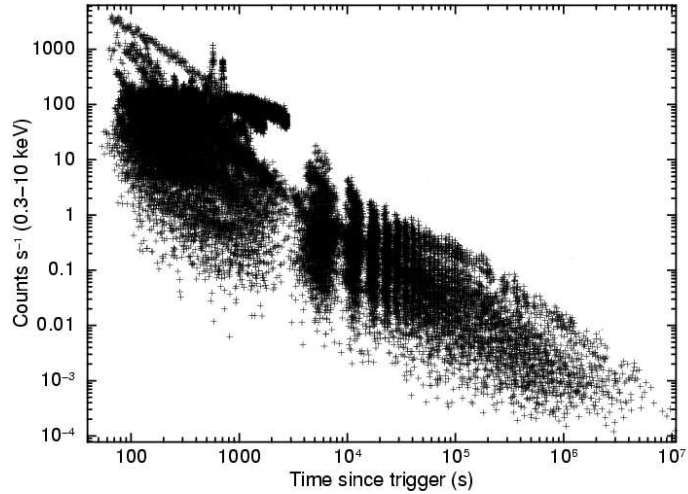}
  \caption{The {\it Swift} XRT light curves for 321 GRBs detected up
  until the early June 2008. The apparent vertical stripe pattern in
  the middle of the plot is due to orbit gaps.}
\label{alllc}
\end{figure}

The X-ray light curves for 321 GRBs detected by the {\it Swift} XRT up
until early June 2008 are shown in Figure~\ref{alllc}. The sheer
number of data points makes it difficult to pick out individual
bursts(!), but a few bright ones can be identified, including the very
bright burst GRB~080319B (the highest data from $\sim 70$s
\citep{racusin08}) and GRB~060124 (the peaks at $\sim 600$ and
700s \citep{ramano06}).  The data illustrate the dynamic range of the
{\it Swift} XRT, covering approximately eight orders of magnitude in
count rate compared to five in time. This illustrates in a simple way
that GRBs decay at faster than $t^{-1}$ on average, even allowing for
the extended X-ray plateau phase. Comparing the observations at 100s
to those at 1 day, the X-ray flux decays approximately as $t^{-1.2}$,
but the path followed varies enormously. The initial range in XRT
count rate (or X-ray flux) is over four decades, similar to the range
in optical flux \citep{kann08}, although the detailed shape of the
light curves are usually different between the optical and X-ray
bands.

\begin{figure}
  \includegraphics[angle=0,width=.4\textheight]
  {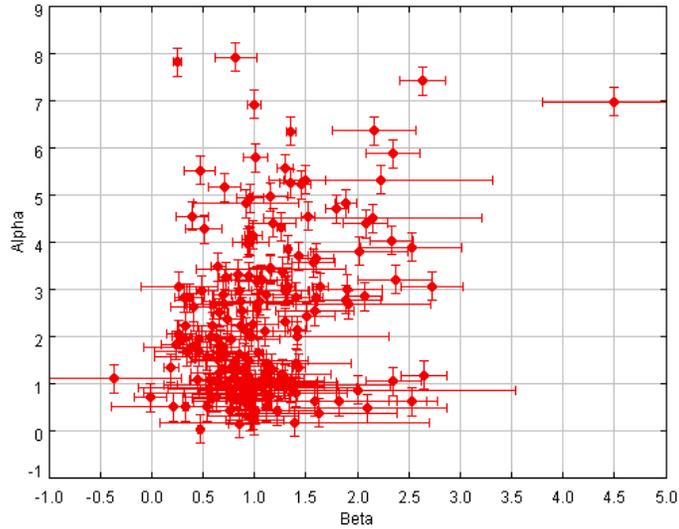} \caption{The temporal (alpha)
  vs. spectral (beta) indices for the early X-ray light curves for
  GRBs observed by {\it Swift}.}
\label{alphabeta}
\end{figure}

The diversity in observed X-ray light curves implies multiple emission
processes contribute to the X-ray emission. Using a functional form with
flux $\propto t^{-\alpha} \nu^{-\beta}$, where alpha and beta are the
temporal and spectral indices respectively, the diversity in the early
X-ray decay and spectral indices is shown in Figure~\ref{alphabeta}.
These data describe the X-ray properties, excluding the effect of
strong flares, during the initial X-ray decay.  The initial X-ray
decay may be due to high-latitude (curvature or off-axis) prompt
emission \citep{ku00} combined with gradually decaying afterglow
emission produced by an external shock \citep{me97}. No single model,
such as the predicted $\alpha = 2+\beta$ from high-latitude emission,
can explain the spread in indices seen in Figure~\ref{alphabeta}.

Emission from an external shock may also explain the plateau phase,
but with additional energy injection to explain the slow decay
\citep{nou06}. A fairly standard afterglow model without much, if any, 
energy injection may explain the entire X-ray decay (except flares)
for those GRBs which show a fairly continuous slow decay from the
earliest times and which have no plateau phase.
A small subset of GRBs in Figure~\ref{alphabeta} have quite a soft X-ray
spectrum yet a modest decay rate (small alpha and high beta). These
are GRBs in which thermal emission may contribute
significantly.

\begin{figure}
  \includegraphics[angle=0,width=.4\textheight]{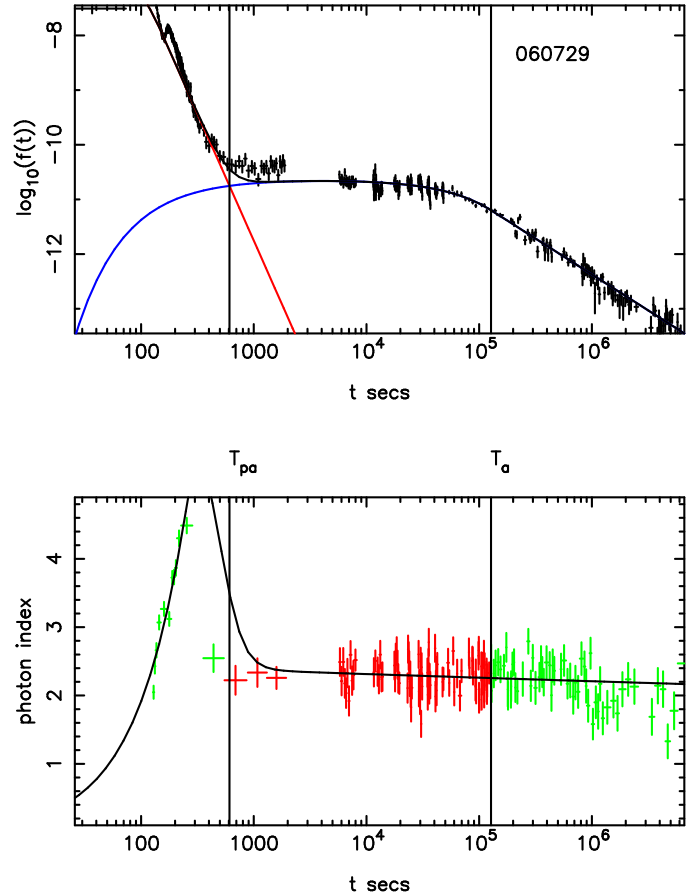}
  \caption{The flux and spectral (photon index) light curves for
  GRB~060729. The fits to the light curves use the methods explained
  in the text. The vertical lines in the lower plot labelled T$_{pa}$
  and T$_{a}$ correspond to the transition from the initial decay to
  the plateau and from the plateau to the final decay respectively. In
  this case, strong spectral softening occurs during the initial decay
  whereas a harder spectrum dominates during the plateau phase and
  final decay.}
\label{lc}
\end{figure}

\section{X-ray spectral variability}

In earlier work {\citep{ob06, will07} we found that the
X-ray light curve can be described well using two generic components
each consisting of an exponential followed by a power-law decay.  The
two components, hereafter called prompt (p) and afterglow (a) dominate
at different times.  The first component (p) usually dominates until
close to the end of the initial steep decline while the second (a)
dominates during the plateau and final decay with an additional late,
break, possibly a jet-break, in some cases \citep{will07}. We proposed
that the p component is the decaying prompt emission while the a
component is emission from the external shock including any energy
injection.

Constraining what physical process dominates when requires additional
information which we can obtain from the observed X-ray spectral
variability. Here we derive spectral information in the form of
time-dependent power-law photon indices ($=\beta+1$) by using the
best-fit late-time absorption column (including Galactic and intrinsic
absorption) combined with hardness ratio data calculated from the
observed counts in the soft (0.3--1.5~keV) and hard (1.5--10.0~keV)
bands \citep{ev07}. We include X-ray counts detected when the XRT was
in Windowed Timing (WT) and Photon Counting (PC) modes. The count-rate
and hardness-ratio data are available at
\url{http://www.swift.ac.uk/xrt_curves/}. 
Example temporal and spectral light curves for GRB~060729
\citep{grupe07} are shown in Figure~\ref{lc}.

A wide range of spectral variability is observed among the GRB
population, but some general trends can be identified which correlate
with the behaviour of the flux light curves. GRBs with steep early
decays tend to get spectrally softer initially before getting harder
and more constant during the X-ray plateau phase and beyond. Other GRBs
show relatively little spectral variability, particularly those with
gradual temporal decays from the start.  The analysis is complicated
in those GRBs with strong X-rays flares as the spectrum tends to get
harder during the flares. Similar conclusions were reached by
\citep{but07, zhang07}, but here we have extended the spectral
analysis to fit across the entire temporal range so we can compare the
initial decay and plateau phase in detail.

Other than during strong flares, where the X-ray spectrum initially
hardens and then softens, those GRBs where strong spectral variability
occurs show spectral softening during the initial decline. In this
short paper we concentrate on those GRBs whose spectra soften
initially, do not have strong flares and which have well determined
plateau phases.  GRB~060729 is a good example of a GRB which shows
early spectral softening. It has a longer than average X-ray
plateau. As can be seen in Figure~\ref{lc}, the spectrum is actually
significantly harder at the start of the plateau compared to the end
of the initial decay, a common pattern in these GRBs, arguing against
a single emission component being responsible for both phases.

To try and model the spectral variability we adapt our previous models
for the temporal behaviour \citep{ob06, will07} but now allow two
components (prompt and afterglow) to vary both in relative strength
{\it and} spectral shape as power-law functions of time. Thus, for the
prompt component the flux varies as $t^{-\alpha_p}$ while the photon
index varies as $t^{-\alpha_h}$. The exact spectral shape of the
later (afterglow) component is not well constrained initially as the
emission is dominated by the prompt component until close to the end
of the initial decay. For simplicity we assume the afterglow component
starts to rise when the prompt component starts to decay. Adopting a
constant flux or spectral shape initially would not significantly
alter the results.  The model predicts that the photon index evolution
illustrated in Figure~\ref{lc} is created by the combination of a
rapidly softening component followed by the emergence of a harder
plateau component.

\begin{figure}
  \includegraphics[angle=-90,width=.4\textheight]
  {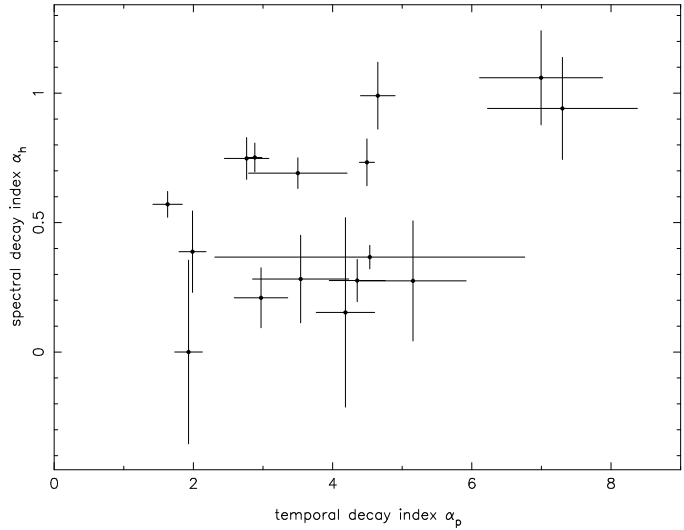}
\caption{The relation between the spectral and temporal decay
indices (defined in the text) for those GRBs with strong initial
spectral softening.}
\label{trend}
\end{figure}

The relation between $\alpha_h$ and $\alpha_p$ is shown in
Figure~\ref{trend} for those GRBs with particularly well determined
values. A correlation exists such that GRBs with the fastest spectral
variability also show the most rapid decline in flux. The most natural
explanation is that a spectral break passes through the band, quite
rapidly, causing a spectral softening and decline in flux (see also
\citep{zhang07}).  The rapidity of the spectral evolution is, however,
difficult to understand. In addition, a single power law usually
provides a good fit to the X-ray data. This also limits the
possibility of additional spectral components, such as thermal
emission, explaining the spectral variability.

\begin{figure}
  \includegraphics[angle=-90,width=.4\textheight]
  {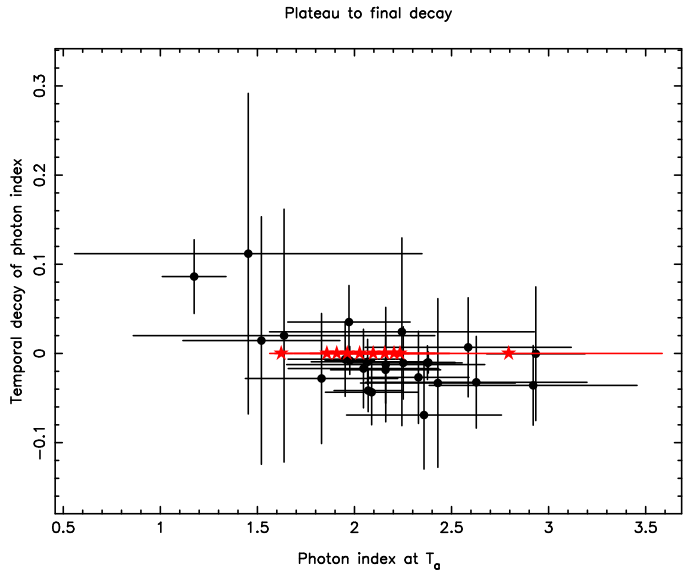}
\caption{The relation between the temporal decay of the photon index
(or spectral index) during the X-ray plateau and the photon index at
the end of the plateau.}
\label{plateauevolve}
\end{figure}

In stark contrast to the initial decline, during the plateau phase,
and subsequently, little or no spectral variability is observed. In
Figure~\ref{plateauevolve} we show the photon-index temporal-decay
rate during the plateau vs. the photon index at the end of the plateau. The
stars show those GRBs for which the data are consistent with no spectral
evolution in the plateau and final decay. The range in spectral decay
indices is much smaller during the plateau than during the initial
decay ($\sim 0.2$ compared to $\sim 1$). Any model put forward to
explains the plateau has to explain this lack of strong spectral
evolution. One possible process is the external shock, which would
require energy injection as noted above for the plateau phase. In some
GRBs the final decay itself is quite slow so exactly when energy
injection ends is unclear \citep{will07}.

\section{Is the X-ray plateau due to dust scattering?}

As an example of using spectral variability to test proposed models,
we have examined the case of an X-ray dust echo origin for the X-ray
plateau. An extended shell of dust in the GRB host galaxy, between the
GRB and observer, would scatter prompt X-rays into the line of sight.
Due to light travel-time effects, these scattered photons would be
observed later and hence could explain the X-ray plateau and
subsequent decay (but not a jet break) seen in many GRBs
\citep{shao07}. This model, however, also predicts strong spectral
evolution across the plateau as the scattering optical depth has an
angular dependence which is a function of energy.

Recently we have compared the predictions of the dust scattering model
with both the temporal and spectral light curves of GRBs
\citep{shen08}. While we confirm that the temporal behaviour of the
X-ray emission can be very well reproduced, the predicted spectral
variability is not seen. This is illustrated in
Figure~\ref{dustscattering} where we show the X-ray light curves for
GRB~060729 in two different energy bands. The X-ray plateau begins and
ends at the same time for each band, whereas the dust echo model
predicts the high-energy plateau should end more than an order of
magnitude earlier in time.

\section{Discussion and conclusions}

The temporal and spectral variability of the GRBs observed using the
{\it Swift} satellite is intriguing. All GRBs show a decaying light
curve, such that by the time a day has passed the X-ray flux has
typically declined by around five orders of magnitude. The exact decay
path, however, varies enormously. The initial decline occurs at
different rates among GRBs and they display different strength X-ray
plateaus. The X-ray plateau is not observed in a significant minority
of GRBs which rather decay gradually throughout. The large majority
of X-ray flares occur before the plateau phase dominates, although 
some are seen at later times.

In GRBs where strong X-ray spectral variability is seen, there is a
correlation between the flux and spectral temporal decay indices. This
is at odds with the simplest expectation of the high-latitude emission
model and may be due to the rapid passage of a spectral break through
the X-ray band. The spectrum hardens at the start of the plateau
followed by relatively little variability, strongly supporting the
idea of a separate origin for the plateau and final decay. A potential
candidate explanation for the plateau and final decay is emission from
an external shock with energy injection --- essentially an afterglow
component \citep{nou06, will07}. Allowing for evolution in the
spectral energy distribution, this may also explain the contrasting
behaviour of the optical and X-ray emission in which most of the
optical emission, except (prompt) flares, may be connected to the
second (afterglow) X-ray component (e.g. \citep{page07, star08}).

Those GRBs which show a fairly uniform flux decay have no evidence for
strong spectral variability, implying their X-ray emission is
dominated by a single process. This may also be the same process ---
the external shock --- as for the plateau and final decay in the
majority of GRBs, but with a much less dominant, if any, contribution
from energy injection. What is as yet unclear is why these GRBs are
distinct from those with a strong plateau phase. 

Recently it has been proposed \citep{kumar08} that fall-back of the
stellar envelope and accretion could explain the shape of the early
X-ray light curve rather than a significant contribution from an
external shock, particularly in those GRBs with significant
plateaus. It is not obvious why such a model would produce the
observed spectral evolution, particularly during the transition from
the initial decay to the plateau or such a constant spectral shape
during the plateau, nor explain the relation with the optical
emission, but a more detailed comparison is required.

\begin{figure}
  \includegraphics[angle=-90,width=.4\textheight]
  {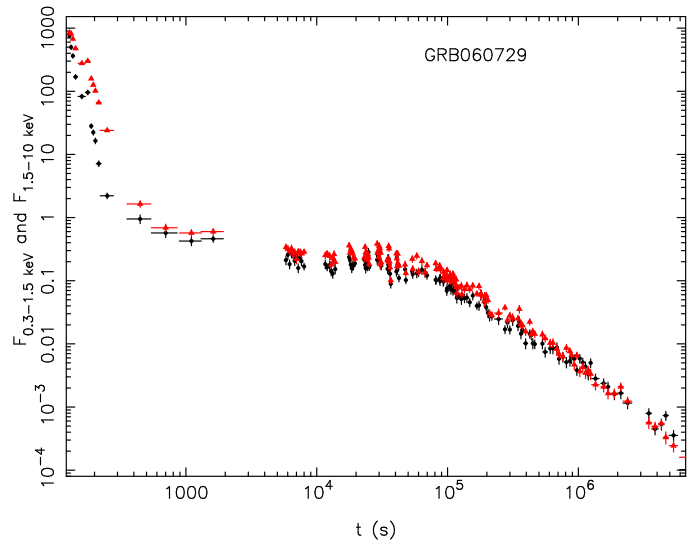}
\caption{The X-ray light curve of GRB~060729 shown in two energy bands:
0.3--1.5~keV (circles) and 1.5--10~keV (triangles). The dust
scattering model predicts the higher energy plateau should have
ended an order of magnitude earlier in time than that of the softer
band, which is clearly not the case.}
\label{dustscattering}
\end{figure}


\begin{theacknowledgments}
The authors gratefully acknowledge funding for {\it Swift} activities
in the UK by the STFC. We also acknowledge useful discussions with our
colleagues on the {\it Swift} team and thank Phil Evans for preparing
Figure 1.
\end{theacknowledgments}



\bibliographystyle{aipprocl} 


\begin{thebibliography}{widest-label}


\bibitem{ge04} N.~Gehrels et al., \emph{ApJ} \textbf{611}, 1005--1020
(2004)

\bibitem{bur05} D.N.~Burrows et al., \emph{Sp. Sci. Rev.} \textbf{120}, 
165--195 (2005)

\bibitem{bart05} S.D.~Barthelmy et al., \emph{Sp. Sci. Rev.} \textbf{120}, 
143--164 (2005)

\bibitem{ta05} G.~Tagliaferri et al., \emph{Nature} \textbf{436}, 
985--988 (2005)

\bibitem{nou06} J.~Nousek et al., \emph{ApJ} \textbf{642}, 389--400 (2006)

\bibitem{ob06} P.T.~O'Brien et al., \emph{ApJ} \textbf{647}, 1213--1237 (2006)

\bibitem{will07} R.~Willingale et al., \emph{ApJ} \textbf{662}, 1093--1110 
(2007)

\bibitem{chin07} G.~Chincarini et al., \emph{ApJ} \textbf{671}, 1903--1920

\bibitem{falcone07} A.D.~Falcone et al., \emph{ApJ} \textbf{671} 1921--1938

\bibitem{kann08} D.A.~Kann et al., \emph{ApJ}, in press (2008)

\bibitem{ku00} P.~Kumar, and A.~Panaitescu, \emph{ApJ} \textbf{541}, 
L51--L54 (2000)

\bibitem{me97} P.~M\'esz\'aros, and M.J.~Rees, \emph{ApJ} \textbf{476}, 
232--237 (1997)

\bibitem{mun07} C.~Mundell et al., \emph{ApJ} \textbf{660}, 
489--495 (2007)

\bibitem{racusin08} J.L.~Racusin et al., \emph{Nature} in press (2008)

\bibitem{ramano06} P.~Ramano et al., \emph{A\&A} \textbf{456}, 
917--927 (2006)

\bibitem{ev07} P.~Evans et al., \emph{A\&A} \textbf{469}, 379--385 (2007) 

\bibitem{grupe07} D. Grupe et al., \emph{ApJ} \textbf{662}, 443--458
(2008) 


\bibitem{but07} N.R.~Butler, and D.~Kocevski, \emph{ApJ} \textbf{668}, 
400--408 (2007)

\bibitem{zhang07} B.-B.~Zhang, E.-W.~Liang, and B.~Zhang, 
\emph{ApJ} \textbf{666}, 1002--1011 (2007)

\bibitem{shao07} L.~Shao and Z.~Dai,
\emph{ApJ} \textbf{660}, 1319--1325 (2007)

\bibitem{shen08} R.F.~Shen et al.
\emph{MNRAS} in press (2008)

\bibitem{page07} K.~Page et al., \emph{ApJ} \textbf{663}, 1125--1138 (2007)

\bibitem{star08} R. Starling et al., \emph{MNRAS} \textbf{384}, 504--514
(2008) 

\bibitem{kumar08} P.~Kumar, R.~Narayan and J.L.~Johnson, 
\emph{Science} \textbf{321}, 376--379 (2008)


\end{thebibliography}


\end{document}